\documentclass[9pt,twocolumn,twoside]{opticajnl}
\journal{opticajournal} 

\setboolean{shortarticle}{false}

\usepackage{lineno}

\newcommand{\de}[0] {{\rm d}}

\title{Passive superresolution imaging of incoherent objects}

\author[1]{Jernej Frank}
\author[1]{Alexander Duplinskiy}
\author[1]{Kaden Bearne}
\author[1,*]{A. I. Lvovsky}

\affil[1]{Department of Physics, University of Oxford, Oxford, OX1 3PU, UK}
\affil[*]{alex.lvovsky@physics.ox.ac.uk}

\begin{abstract}
We investigate Hermite Gaussian Imaging (HGI) --- a novel passive super-resolution technique --- for complex 2D incoherent objects in the sub-Rayleigh regime. The method consists of measuring the field's spatial mode components in the image plane in the overcomplete basis of Hermite-Gaussian modes and their superpositions and subsequently using a deep neural network to reconstruct the object from these measurements. We show a three-fold resolution improvement over direct imaging. Our HGI reconstruction retains its superiority even if the same neural network is applied to improve the resolution of direct imaging. This superiority is also preserved in the presence of shot noise. Our findings are the first step towards passive super-resolution imaging protocols in fluorescent microscopy and astronomy.
\end{abstract}

\setboolean{displaycopyright}{false} 

\begin{document}

\maketitle

\section{Introduction}
With the human eye, we can only see objects as small as a thin strand of hair, which is about 100 microns in width \cite{yanoff2008ophthalmology}. Modern life and materials sciences are keen to study structures well beyond this limit, and microscopy is an indispensable tool for this purpose. The resolution of conventional optical microscopy is, however, limited due to diffraction on the objective lens to approximately a half of the optical wavelength ($\sim$200 nm) \cite{rayleigh1879xxxi,abbe1873beitrage,abbe1883xv}. In the last few decades, super-resolution (SR) imaging techniques emerged focusing on either manipulating the illumination of the sample or using stochastic properties of fluorescent emitters to achieve resolution well beyond the diffraction limit \cite{pujals2019super,huszka2019super,valli2021seeing,liu2022super}. Some of these methods were marked by the 2014 Nobel Prize in chemistry \cite{mockl2014super}. Even higher resolution --- down to single nanometers --- can be achieved by atomic force microscopy \cite{binnig1986atomic} and electron microscopy \cite{mathys2004entwicklung}. However, all existing superresolution methods involve active interaction with the sample, which imposes significant limitations. First, they involve the risk of altering or damaging the samples. Second, these techniques are not universally applicable to all imaging scenarios since they rely on specific underlying properties of the samples --- such as, e.g.,~nonlinear susceptibility. Even if the sample is amenable to a particular method, SR microscopes require expert knowledge, and various parameters must be adjusted on a case-by-case basis to render a truthful image reconstruction.  

These limitations can be overcome if the imaging method involves only \textit{far-field, linear-optical, and passive measurement of the optical field arriving from the object}. In 2016 Tsang {\it et al.} \cite{tsang2016quantum, tsang2017subdiffraction, tsang2019resolving, tsang2019resurgence, tsang2019quantum}
theoretically predicted that the diffraction limit can be overcome for estimating the distance between two incoherent point sources in the far-field regime by decomposing the field in the image plane into an orthonormal basis of spatial modes (typically, Hermite-Gaussian) and measuring the intensity of each component. This approach, dubbed SpaDe for ``spatial demultiplexing''  was soon experimentally demonstrated by various groups \cite{yang2016far,tang2016fault, paur2016achieving, tham2017beating, parniak2018beating,zanforlin2022optical,ozer2022reconfigurable}. 

On this basis, Yang {\it et al.} proposed an extension into reconstructing complete 2D objects dubbed Hermite-Gaussian Imaging (HGI) \cite{yang2016far}. Pushkina {\it et al.} experimentally implemented HGI for coherent 2D objects and showed a two-fold resolution improvement over the diffraction limit \cite{pushkina2021superresolution}. 

However, the latter experimental work used coherent light to illuminate the sample and detect the reflected light. On the other hand, fluorescence emitted by biological samples is incoherent. Also, if SpaDe is to be applied outside microscopy --- say, in astronomy or remote imaging, it needs to be adapted to incoherent input. Here we address this requirement by extending HGI to imaging incoherent 2D objects. Since the incoherent light emitted from the object lacks phase information, it is necessary to measure in an overcomplete spatial mode basis, involving not only HG modes but also their superpositions. We measure the intensity in each basis mode using heterodyne detection and then utilize a deep neural network (DNN) to reconstruct the objects. We observe a significant resolution improvement over direct imaging (DI) in various scenarios and certified by quantitative benchmarks.

\section{Theoretical Background}

\subsection{Image reconstruction theory}
We briefly recap the reconstruction method of Yang {\it et al.} \cite{yang2016far}. The object being imaged is defined by the incoherent intensity distribution $I(x,y)$. We reconstruct this distribution by expanding it into the HG basis
\begin{align}
    I_{\text{rec}}(x,y) = \sum_{m,n = 0}^N \frac{\beta_{mn} \phi_{mn}(x,y)}{2^{m+n+2}m!n!\pi\sigma^2},
    \label{eq:img_rec}
\end{align}
where $\phi_{mn}(x,y) = H_m\left(\frac{x}{\sqrt{2}\sigma}\right)H_n\left(\frac{y}{\sqrt{2}\sigma}\right) e^{-\frac{x^2+y^2}{4\sigma^2}}$ are the HG polynomials, $N$ is the cutoff for the number of HG modes to be used, $\sigma \approx 0.21 \lambda / \text{NA}$ is the radius of the PSF at $e^{-1/2}$ intensity \cite{zhang2007gaussian}, and $\beta_{mn}$ are the expansion coefficients. The PSF mode is assumed to match $\phi_{00}$. 

\begin{figure}[H]
    \centering
    \includegraphics[width=\linewidth]{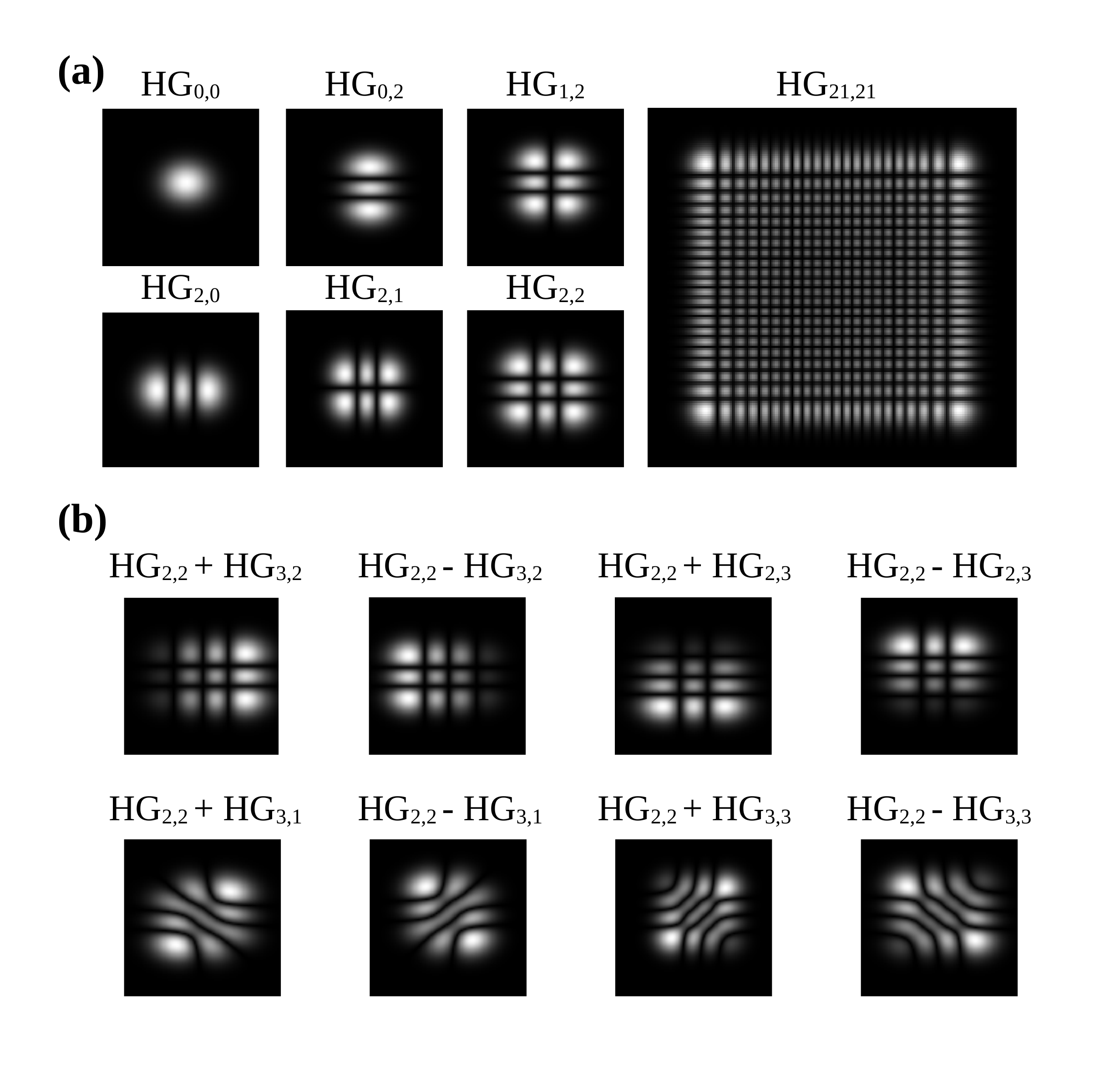}
    \caption{\textbf{(a)} Subset of holograms displayed on the SLM to generate the LO in Hermite-Gaussian modes. \textbf{(b)} Expanded iHG basis, which includes 8 additional superpositions of neighbouring HG modes to obtain the odd geometric moments for image reconstruction.}
    \label{fig:ihg_basis}
\end{figure}

These coefficients can be obtained experimentally by measuring the powers of the heterodyne detector photocurrents with the local oscillator prepared in a set of modes (see Appendix for a full derivation). This set of modes can be the HG basis [Fig.~\ref{fig:ihg_basis}(a)]; however, in this case, we are only able to recover the $\beta$'s with even indices, which, in turn, allow us to reconstruct only the even component of the object intensity distribution:  $I_{\rm even}(x,y)=[I(x,y)+(-x,y)+I(x,-y)+I(-x,-y)]/4$, which results in ghost image artefacts \cite{yang2016far,tsang2017subdiffraction}. To address this problem, Yang {\it et al.}\cite{yang2016far} proposed displacing the object into a single quadrant of the x-y plane, thereby separating the image from its ``ghosts''. However, the displaced object together with the ``ghosts'' is larger in size and requires higher-order modes for its expansion into the HG basis. Furthermore, this approach is unsuitable for objects with a wide spatial extent.

Tsang {\it et al.} proposed an alternative strategy: recovering the odd components of the image by measuring the powers in an overcomplete mode basis known as interferometric HG (iHG) \cite{tsang2017subdiffraction,tsang2018subdiffraction}. The version of the iHG basis used in our work includes the following superpositions of individual HG modes [Fig. \ref{fig:ihg_basis}(b)]:
$$\phi_{mn}^{0-8}:=\{\phi_{mn}, \phi^\pm_{m,m+1, n,n}, \phi^\pm_{m,m,n,n+1}, \phi^\pm_{m,m+1,n,n+1} , \phi^\pm_{m,m+1,n,n-1} \},$$
where
\begin{align}
    \phi^\pm_{m,m',n,n'} := \frac{1}{\sqrt{2}} [\phi_{m'n'} \pm \phi_{mn}].
\end{align}
For each mode $\phi_{mn}^{i}$, the power $P_{mn}^{i}$ of the associated heterodyne photocurrent is measured. These measurements then evaluate the coefficients $\beta_{mn}$, from which the image is recovered (see Appendix). Note that theoretically, iHG modes $\phi_{mn}^{i}$ with $i\in\{0,\ldots,6\}$ are sufficient for complete reconstruction. However, the experiment uses the entire iHG mode set with $i\in\{0,\ldots,8\}$, as defined above.

\subsection{Simulations}
\label{sec:numerical_analysis}
To test the viability of our approach we first simulate image reconstruction computationally using \eqref{eq:img_rec} and compare it to simulated DI. We set the numerical aperture (NA) to $5.5 \times 10^{-4}$ and $\lambda = $  785 nm (the same as used in the experiment).

\begin{figure}[H]
    \centering
    \includegraphics[width=\linewidth]{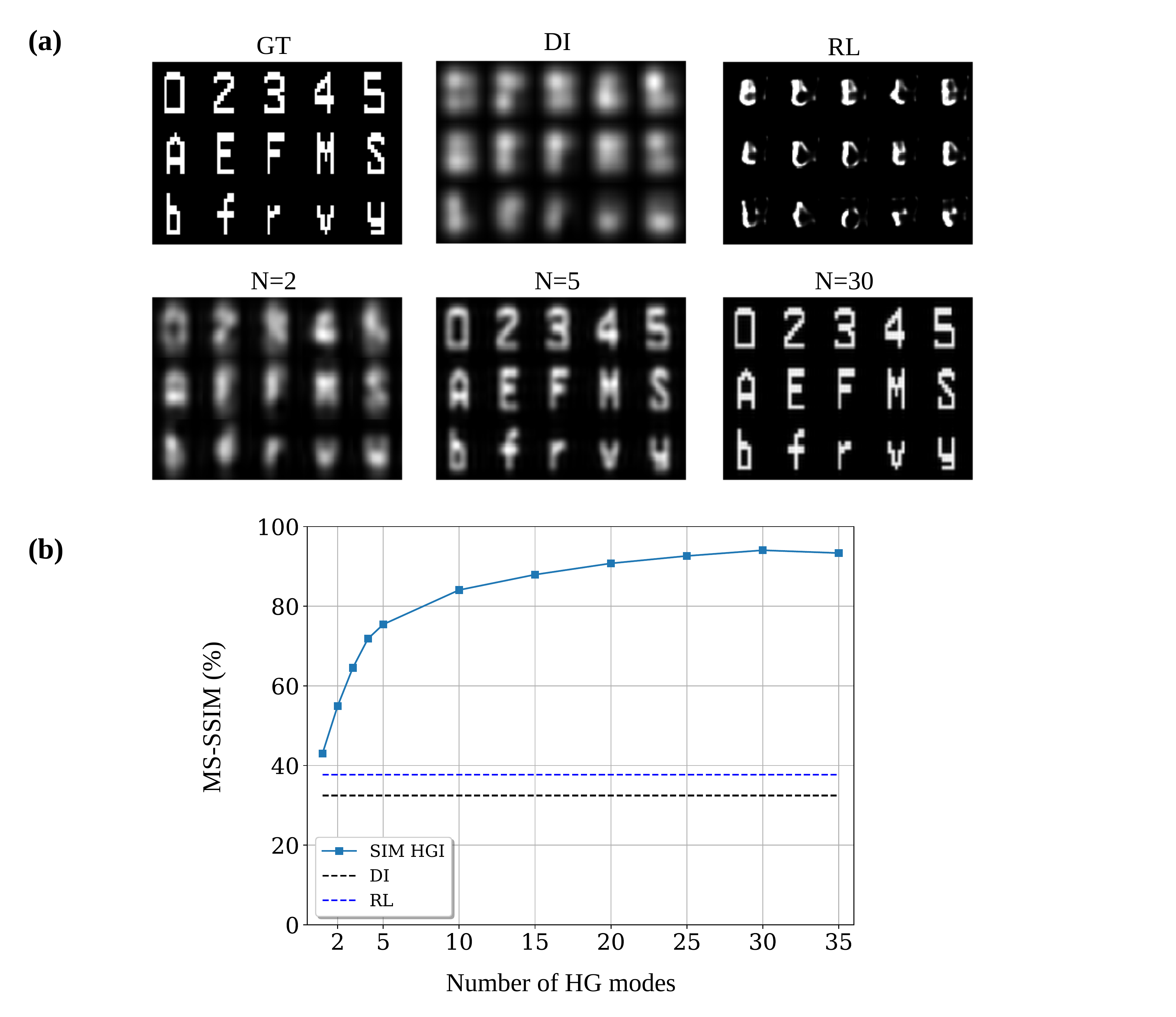}
    \caption{\textbf{(a)} Letters of the alphabet with varying resolution. Top row: GT is the ground truth, DI represents the direct image whose resolution is limited by diffraction, RL is the Richardson-Lucy deconvolution of the DI. Bottom row: simulated HGI reconstructions in an idealized setting with up to $m=n=N=2,5,30$ modes. \textbf{(b)} MS-SSIM calculated between (1) the ground truth (infinite resolution) and (2) the simulated HGI, DI, and RL reconstructions. MS-SSIM score of $\gtrsim80\%$ achieves letter readability.}
    \label{fig:sim_alphabet_piqa}
\end{figure}

As our objects, we use symbols of a bitmap ASCII font. Each symbol is a black-and-white bitmap $8\times 6$ pixels, with the size of each pixel being 0.36 $\sigma$~[Fig.~\ref{fig:sim_alphabet_piqa}(a), upper left panel]. To simulate DI, we convolve the PSF with  the objects [Fig.~\ref{fig:sim_alphabet_piqa}(a), upper middle panel]. The upper right panel of Fig.~\ref{fig:sim_alphabet_piqa}(a) shows the result of applying the conventional Richardson-Lucy (RL) deconvolution algorithm \cite{richardson1972bayesian} to the DI data. Next, we simulate the image reconstruction for various cut-off numbers  $N=2,5,30$ [Fig.~\ref{fig:sim_alphabet_piqa}(a), bottom row]. As can be seen, the DI of most of the letters is unreadable, whether or not RL is applied. But HGI allows us to achieve a satisfactory level of readability, the quality improving with $N$. 

We quantify the image quality with respect to the ground truth objects by computing the Multi-Scale Structural Similarity (MS-SSIM) \cite{msssim_wang} [Fig.~\ref{fig:sim_alphabet_piqa}(b)], which is a popular metric in the image processing community, designed to mimic how the human eye perceives images \cite{zhao2016loss}. We find that the resolution improvement is proportional to $N$ and saturates for $N>25$. Comparing the images with the MS-SSIM graph in Fig.~\ref{fig:sim_alphabet_piqa}, we can see that satisfactory image quality is achieved at about $80\%$.

\section{Experiment}
\subsection{Setup}
\begin{figure}[H]
    \centering
    \includegraphics[width=\linewidth]{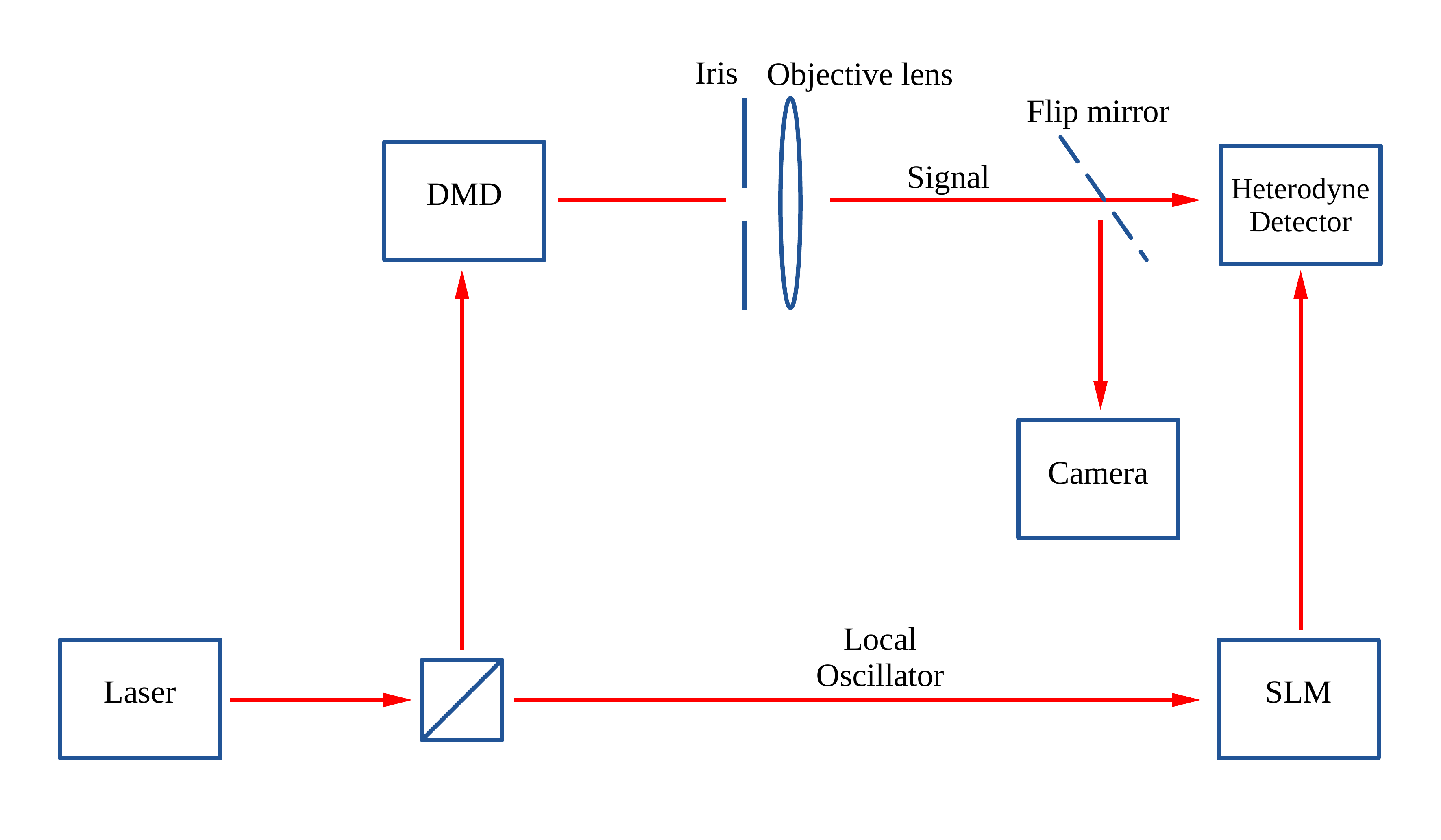}
    \caption{Experimental setup to image incoherent objects with HGI via homodyne detection and DI.}
    \label{fig:exp_setup}
\end{figure}

We adopt the experimental setup from Ref.~\cite{pushkina2021superresolution} (Fig.~\ref{fig:exp_setup}). We use a continuous wave diode laser (Eagleyard EYP-DFB-0785), operating at 785 nm. The powers of the image field in individual HG modes are measured by means of heterodyne detection with the local oscillator (LO) prepared in the corresponding mode. 

At the outset, we split the laser into the signal LO paths. We use the signal path to illuminate the object displayed on a digital micromirror device or DMD (DLP LightCrafter 6500). The DMD pixel pitch is 7.56 $\mu$m, and our objects are displayed in a square area of $160 \times 160$ pixels. This area is divided into $10\times10$ ``logical pixels", each being a square of $16\times16$ physical DMD pixels. Our binary objects are bitmaps of these logical pixels. To simulate incoherent objects, the DMD displays logical pixels one at a time, and for each pixel, photocurrent powers $P_{mn}^{i}$ for all iHG modes are measured. In order to determine $P_{mn}^{i}$ corresponding to a given multi-pixel object, we digitally add the measured values of $P_{mn}^{i}$ for all of the object's pixels. This approach is chosen instead of the usual method of obtaining spatial incoherence via a rotating ground glass because the latter method would require averaging of the speckle pattern for each object and each combination $(m,n,i)$, resulting in prohibitively slow data acquisition.

After the light is reflected from the object we collect it by an objective lens. In front of the objective lens, at a distance 242.5 cm away from the DMD, we place an iris of diameter 2.66 mm  to reduce the NA to $5.5 \times 10^{-4}$. This configuration sets the Rayleigh diffraction limit to a spacing of 7 logical DMD pixels (870 $\mu$m). 

To generate a particular iHG mode in the LO path, we display a phase grating (``hologram'') on a reflective phase-only liquid-crystal-on-silicon SLM (Hamamatsu X13138-02) and select the first diffraction order of the reflected light \cite{bolduc2013exact,pushkina2020comprehensive}. We switch through holograms corresponding to different iHG modes sequentially.

The signal and LO are recombined on a beam splitter and sent onto a balanced detector. 
By placing a flip mirror after the objective lens we are able to switch between recording HGI with the heterodyne detector and DI using a high-resolution CMOS camera.

\subsection{Deep neural networks}
As realized in Ref.~\cite{pushkina2021superresolution}, the experimental imperfections such as an imperfect Gaussian PSF, optical aberrations, shot noise, electronic white noise, optical misalignment, iHG mode fidelity in the LO, imperfect overlap between signal and LO, and general air fluctuation or vibrations render an image reconstruction using \eqref{eq:img_rec} impossible because the measurement errors of lower order iHG modes propagate into higher order $\beta_{mn}$ coefficients. 

We, therefore, recast our reconstruction problem into a machine-learning task and let a DNN find an optimal mapping between imperfect measurements and the underlying basis needed for image reconstruction. In this way, we can benefit from both the new physics of HGI measurement and machine learning to extract the image with the highest possible  resolution. The neural network takes the measured photocurrent powers as input and outputs the reconstructed image.

\begin{figure}[H]
    \centering
    \includegraphics[width=\linewidth]{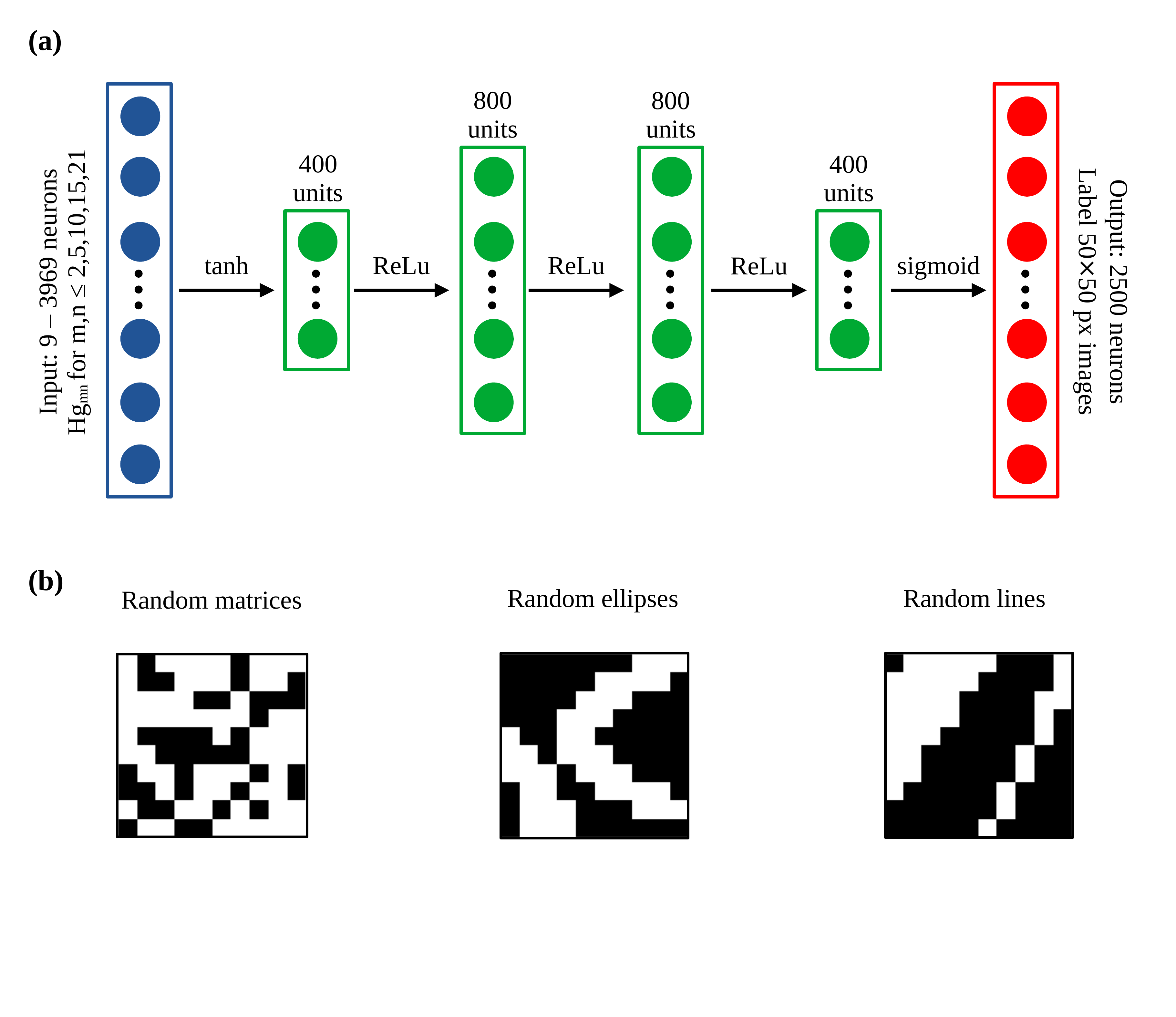}
    \caption{\textbf{(a)} Architecture of the DNN used for image reconstruction from iHG modes. The input layer (blue) dimension varies with the number of iHG modes utilized. The hidden layers (green) reduce the latent space to 400 neurons. The output layer (red) is fixed to output images $50\times 50$ pixels. \textbf{(b)} Examples of bitmaps used for training the DNN.}
    \label{fig:NN}
\end{figure}

The DNN architecture is shown in Fig.~\ref{fig:NN}(a). We have four fully-connected hidden layers with 400-800-800-400 neurons. The activation functions are hyperbolic tangent in the first layer ReLU in the next three layers; the last layer adapts the [0,1] range by using sigmoid.

To acquire the DNN training and cross-validation sets, we obtain the photocurrent powers in iHG modes $\phi_{mn}^l$, with $m,n\in\{0,\ldots,20\}$ and $l\in\{0,\ldots,8\}$ for 20000 bitmaps representing random matrices, ellipses, and lines[Fig.~\ref{fig:NN}(b)]. In order to investigate the dependence of the reconstruction quality on the number of modes acquired, several subsets are selected from the above set, corresponding to $m,n\le M$ with $M=1,4,9,14,20$. These subsets are split into training and cross-validation sets in proportion 90:10. 

Using the ground truth images for labels would lead to severe overfitting \cite{pushkina2021superresolution}. We, therefore, produce our labels according to \eqref{eq:img_rec}. That is, our label is the image of the object as if it were reconstructed via HGI in the absence of imperfections. We produced several label sets of varied resolution by choosing the mode cut-off number $N$ between 5 and 35. 

For the DNN training, we set our loss function to be the mean squared error (MSE) and use a minibatch size of $32$. We use the adaptive moment estimation (Adam) optimiser, learning rate $10^{-4}$, exponential moving averaging parameters for the first and second moment estimates $(0.9, 0.999)$ respectively, and weight decay $0$. The training is run on an 11th Generation Intel Core i9-11900K CPU with 64 GB memory and an NVIDIA® RTX™ A5000 GPU with 24 GB memory. Training the DNN up to $2000$ epochs takes 40--60 minutes on average.

\section{Results}
\label{sec:results}
We evaluate the performance of our method using the alphabet letters [Fig.~\ref{fig:sim_alphabet_piqa}(a)] as well as pairs of parallel lines with varying separation. None of these objects belongs to the training set. The experimental data for these objects have been acquired separately from the training set data. 

We train several DNNs where we vary both the number $M$ of iHG modes in the input and the number $N$ of HG modes used in simulating the labels. To show that the resolution improvement arises due to the new physics of HGI measurements and not only thanks to machine learning we also trained the same DNN  using DI intensity measurements as inputs. These inputs are $63\times 63 = 3969$ pixel bitmaps which correspond to roughly the same dimension of the input layer as for  iHG with $M=20$.

\begin{figure}[H]
    \centering
    \includegraphics[width=\linewidth]{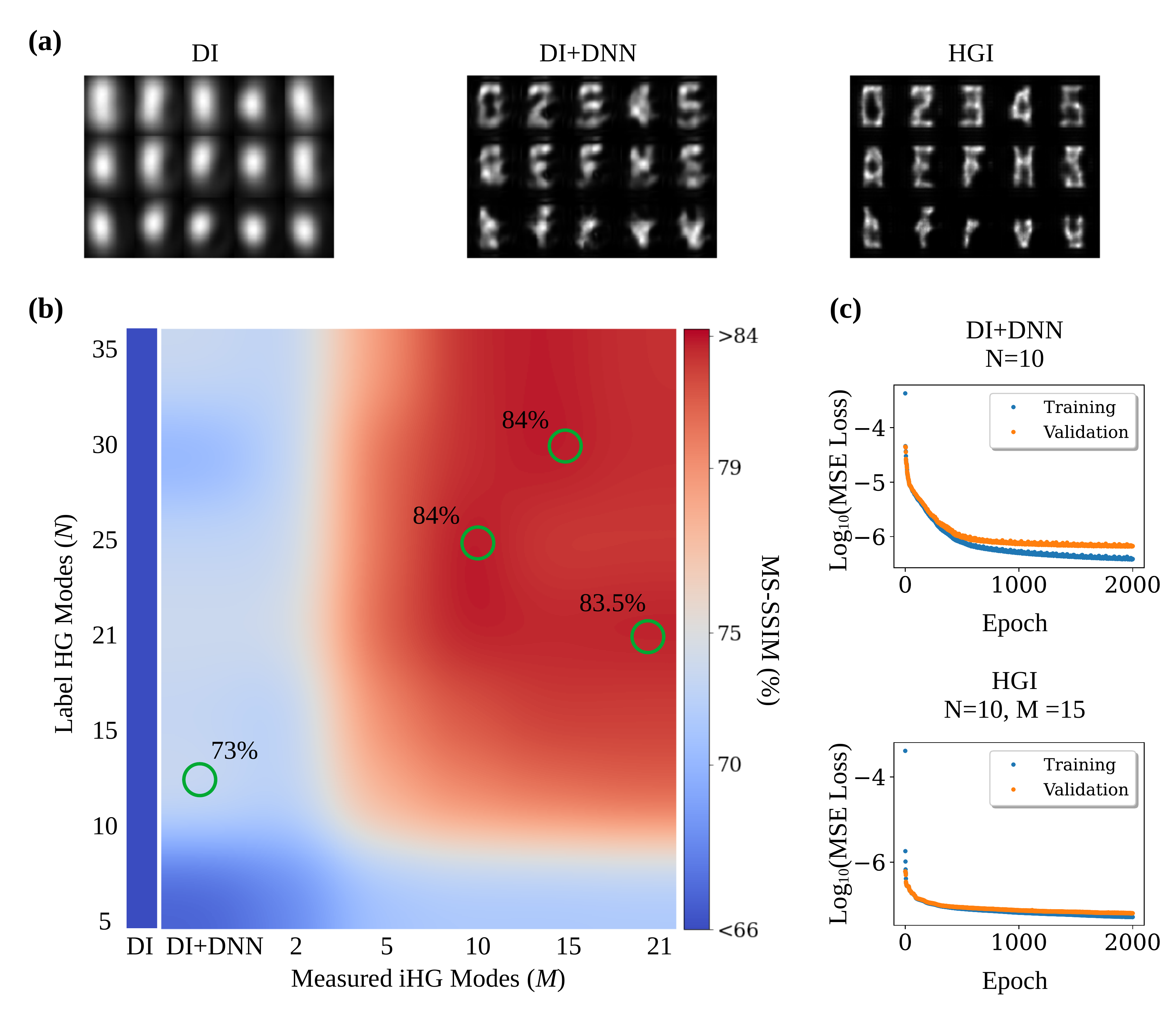}
    \caption{\textbf{(a)} Alphabet letters imaged using DI (left), DNN-enhanced DI (middle), and HGI using $M=15$ iHG modes (right). \textbf{(b)} Multi-scale structural similarity values for different pairs of input and label mode numbers. \textbf{(c)}  Training and cross-validation MSE as a function of training epoch. DI+DNN overfits and we stop training at 1000 epochs, while for HGI we train for the complete 2000 epochs. 
    }
    \label{fig:HGI_density}
\end{figure}

We compare the results qualitatively by looking at the image reconstruction results [Fig.~\ref{fig:HGI_density}(a)] and quantitatively by computing the MS-SSIM between the ground truth (infinite resolution) and the respective outputs of the DNNs [Fig.~\ref{fig:HGI_density}(c)].

We see that the resolution improvement is maximized for $N=30$ label modes and $M=10$ to $15$ measured iHG modes. We attribute the lack of resolution improvement for higher $M$ to the photocurrents in higher-order iHG modes being very low and hence prone to noise.

We can also observe some resolution improvement beyond the diffraction limit with DI purely through DNN software enhancement. However, this enhancement is smaller than that of HGI. We can see this directly from the reconstructed images [Fig. \ref{fig:HGI_density}(a)] and from the MS-SSIM benchmarks [Fig. \ref{fig:HGI_density}(b)]. We also observe that training the DNN for DI with the number $N$ of label modes as low as 10 quickly leads to overfitting, which reduces the reconstruction quality on the test set [Fig. \ref{fig:HGI_density}(c)].

\begin{figure}[H]
    \centering
    \includegraphics[width=\linewidth]{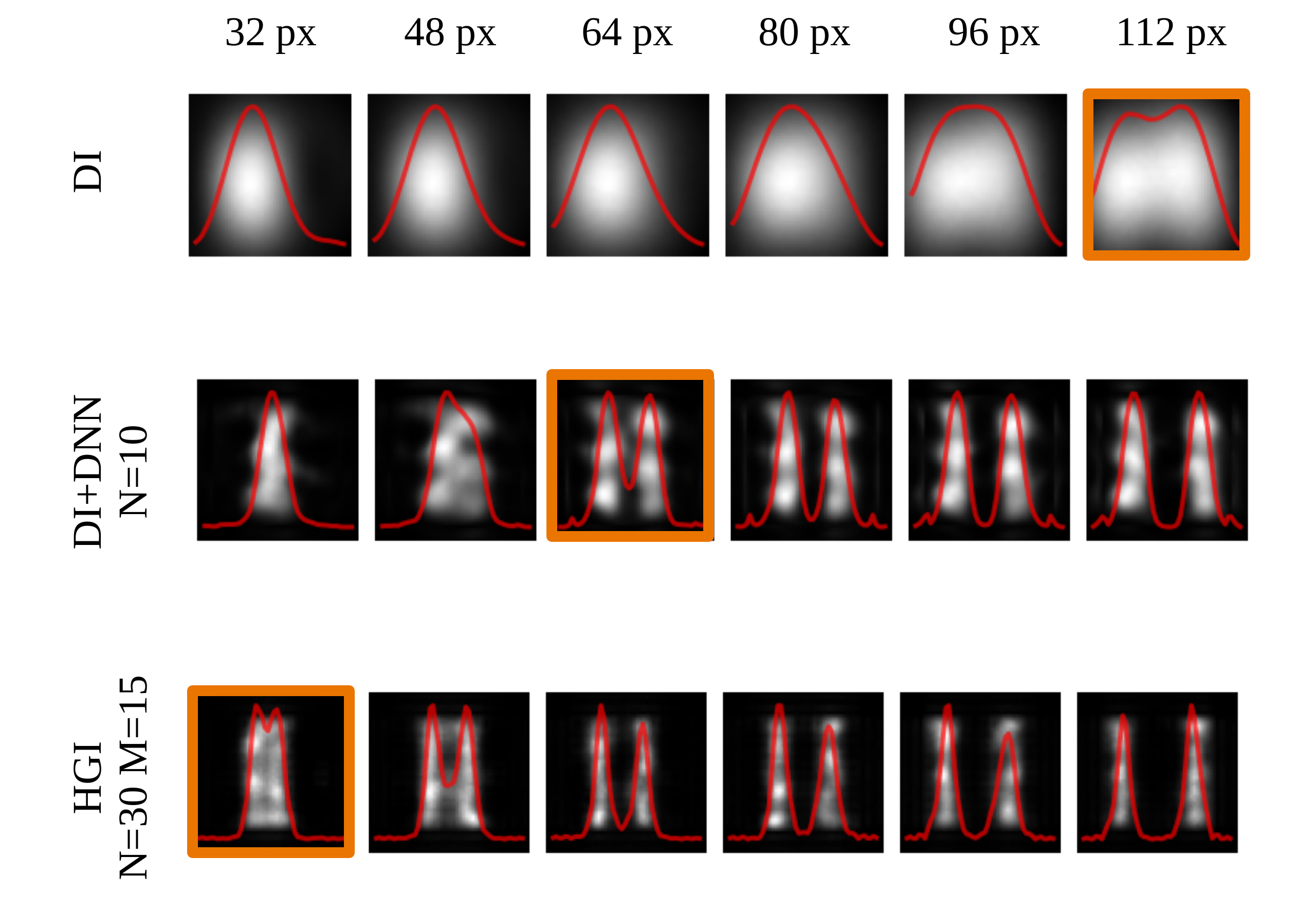}
    \caption{Reconstructed images of pairs of parallel lines via direct imaging (top), DNN-enhanced direct imaging (middle) and HGI (bottom). Coloured frames correspond to the resolution limits in the three settings.}
    \label{fig:rec_RL}
\end{figure}

To determine the resolution limit, we reconstruct the images for pairs of parallel lines and measure the intensity profile by integrating over the vertical dimension [Fig.~\ref{fig:rec_RL}]. We see that the resolution of direct imaging is limited by the diffraction limit, corresponding to a line separation of 7 logical DMD pixels ($847\,\mu$m). DI+DNN gives some improvement over that limit by resolving lines 4 logical pixels ($484\, \mu$m) apart, but HGI is able to resolve the separation of 2 logical pixels ($242\, \mu$m), which compared to DI+DNN still yields a factor of 2 resolution enhancement.

Lastly, we investigate the effect of simulated shot noise on the performance of HGI and DI. To this end, we assume that all measurements (iHG mode intensities in HGI and pixel intensities in DI) are implemented via photon counting, and the total number of photons detected is $N_{\rm ph}$. Then the number of photons $\lambda_i$ in each measurement is $N_i=N_{\rm ph}\lambda_i/\sum_i\lambda_i$ and the corresponding rms shot noise is $\Delta N_i/N_i=\Delta\lambda_i/\lambda_i=1/\sqrt{N_i}$. We simulate the shot noise by adding  a normally distributed random value with the rms deviation $\Delta\lambda_i$ to every $\lambda_i$. 

\begin{figure}[H]
    \centering
    \includegraphics[width=\linewidth]{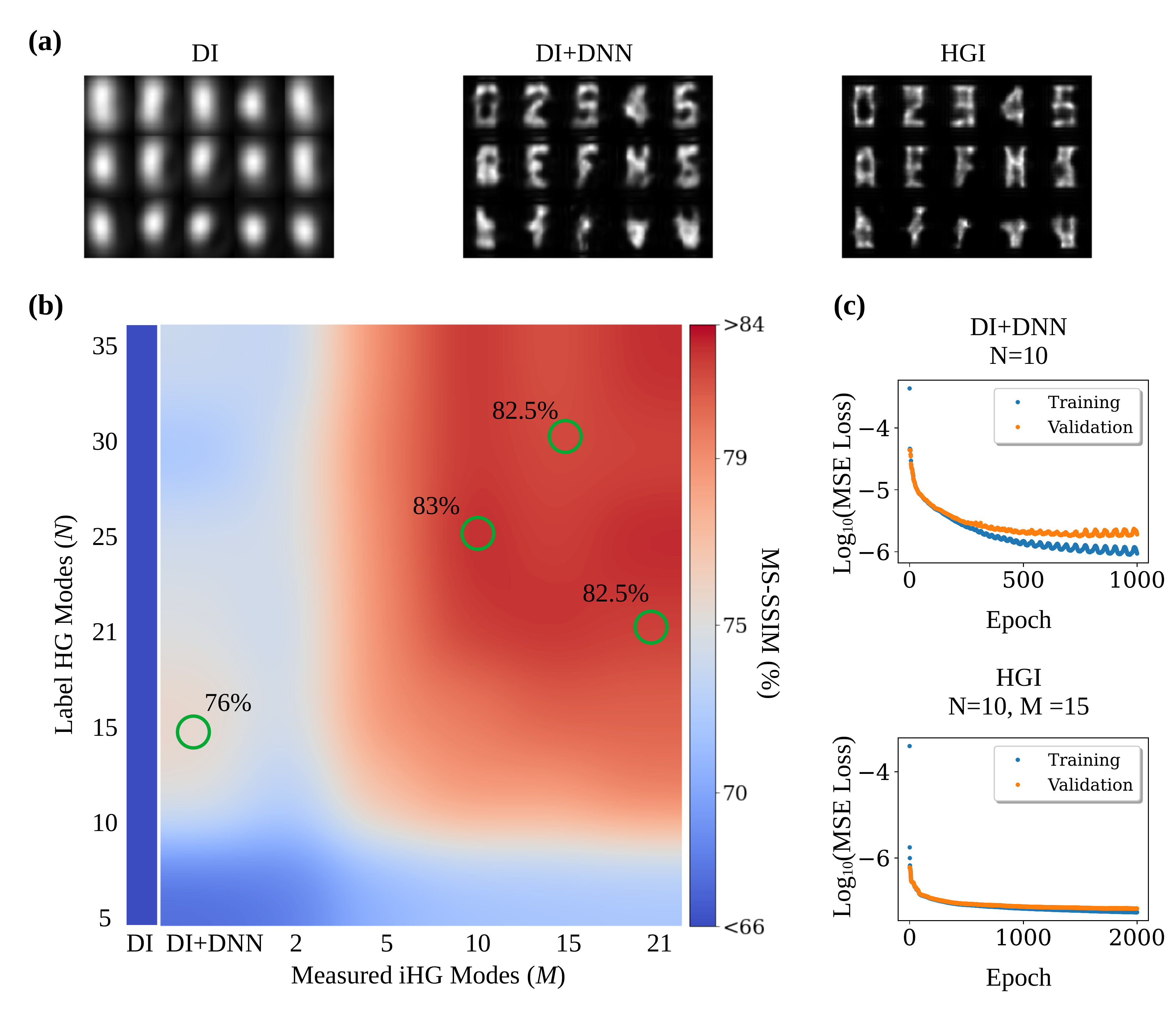}
    \caption{\textbf{(a)} Alphabet letter images with simulated shot noise assuming $N_{\rm ph}=10000$ photons using DI (left), DNN-enhanced DI (middle), and HGI using $M=15$ iHG modes (right). \textbf{(b)} MS-SSIM values for different pairs of inputs and labels. The performance is similar to the classical case. \textbf{(c)} Training and cross-validation MSE as a function of training epochs for selected input and label pairs. DI+DNN overfits at 300 epochs, at which point the training is stopped, while the DNN for HGI is trained for 2000 epochs. }
    \label{fig:HGI_density_noise}
\end{figure}

We set $N_{\rm ph}$ to 100, 400 and 10,000 and add the photon shot noise numerically before training the DNNs. As can be seen from Fig.~\ref{fig:HGI_density_noise}, the advantage of HGI over DI for $N_{\rm ph}=10^4$ is largely similar to the classical case. Particularly strong is the impact of the shot noise on the loss function of DI+DNN [Fig.~\ref{fig:HGI_density_noise}(c)]: the neural network begins overfitting as early as at epoch 300. 

\begin{figure}[H]
    \centering
    \includegraphics[width=\linewidth]{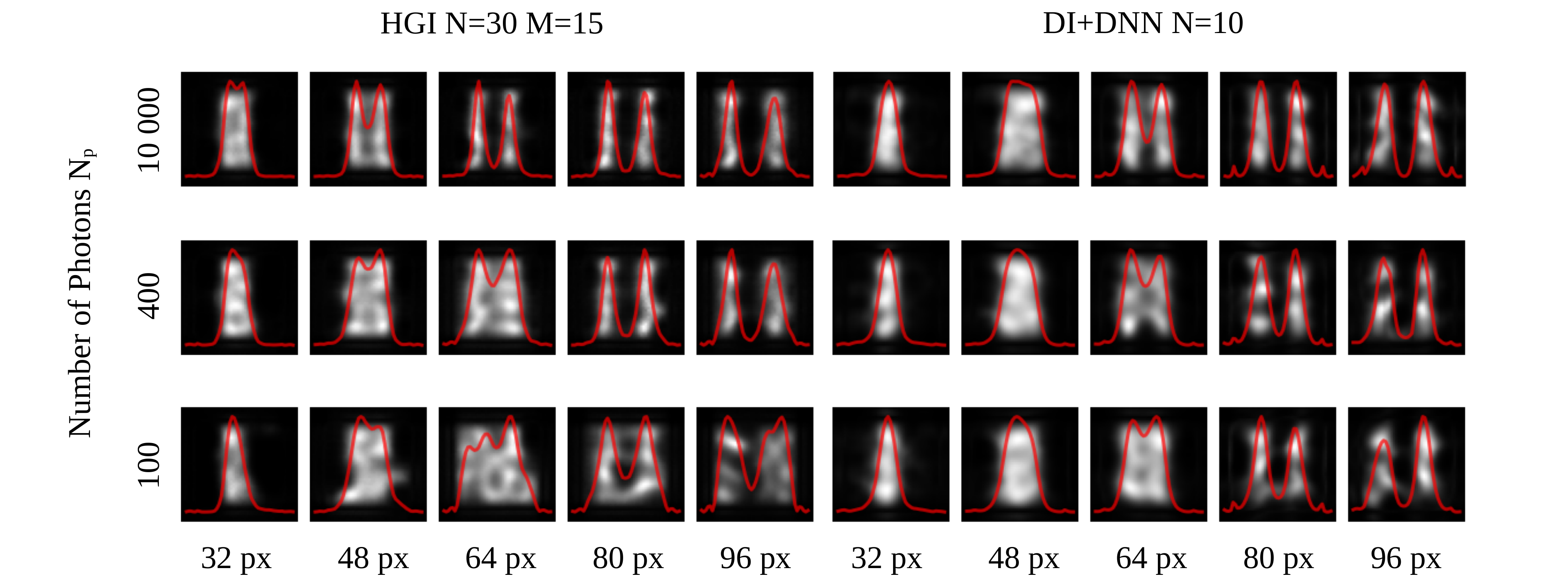}
    \caption{Reconstructed images using HGI (left) and DI+DNN (right) of pairs of parallel lines with added photon shot noise for $N_{\rm ph}=100,400,10.000$.}
    \label{fig:rec_RL_noise}
\end{figure}

On the other hand, the edge of HGI over DI+DNN degrades for $N_{\rm ph}=400$ and especially $100$ [Fig.~\ref{fig:rec_RL_noise}]. This is because the total number of photons available is no longer sufficient to reliably reconstruct the relatively intricate image even with the help of HGI.

\section{Conclusion}
Here we presented the first results in superresolving reconstruction of complex incoherent 2D objects by completely passive measurements using linear optics in the far-field regime. We have shown that a combination of information-rich  measurements with a simple DNN  produces images with $\sim3$-fold resolution improvement over DI. We can enhance the DI resolution by applying the same DNN; however, the resulting resolution is still a factor of $\sim2$ below that achieved with HGI. This shows the key role of the new physics of HGI in resolution enhancement. 

Due to its passive nature, HGI can become beneficial in various scenarios far beyond microscopy. This includes  astronomy, remote sensing, satellite imaging and many other settings where active interaction with the object is impossible. The main obstacles to the practical application of our scheme are the training of the DNN required for the reconstruction and the bandwidth limitations of heterodyne detection. To address the latter concern, the homodyne detector needs to be replaced by a mode sorter --- a device that would spatially separate individual modes of the incoming light field, after which the intensities of these modes can be measured by direct photodetection \cite{fontaine2019}. 

The training set samples can be fabricated using commercially available fluorescent beads in various sizes. Alternatively, calibration slides  can be fabricated with a resolution of a few tens of nanometers through lithography or laser writing. In both cases, the training samples will need to be imaged employing electron microscopy or another method to obtain ground truth data. With the rapid advancement of machine learning technology, we can also envision that better DNN architectures and loss functions will emerge that would drastically reduce the size of the required training set. In the meantime, we can expand the training set using known data augmentation techniques \cite{shorten2019survey}.

It would be interesting to combine HGI with existing passive SR microscopy methods based on engineered sample illumination, such as confocal and structured-illumination microscopy \cite{bearne2021confocal}. Because these methods and HGI rely on different physics to enhance the resolution, we expect the enhancement effect achieved through both methods to be cumulative. In this way, we hope passive microscopy resolution with engineered illumination to eventually reach the scale of 50 nm. 

\begin{backmatter}
\bmsection{Funding} The project is funded by BBSRC grant BB/X004317/1. JF is supported by the European Union’s Horizon 2020 research and innovation programme under the Marie Skłodowska-Curie grant agreement No 956071. 


\bmsection{Disclosures} The authors declare no conflicts of interest.

\bmsection{Data Availability Statement} Data underlying the results presented in this paper are not publicly available at this time but may be obtained from the authors upon reasonable request.

\bmsection{Supplemental document}
See Supplement 1 for supporting content. 

\end{backmatter}

\bibliography{sample}


\appendix
\pagebreak
\onecolumn
\begin{center}
\textbf{\large Appendix: Derivation of Eqs.~(1) and (2)}
\end{center}
\setcounter{equation}{0}
\setcounter{figure}{0}
\setcounter{table}{0}
\setcounter{page}{1}
\makeatletter
\renewcommand{\theequation}{S\arabic{equation}}
\renewcommand{\thefigure}{S\arabic{figure}}
\renewcommand{\bibnumfmt}[1]{[S#1]}
\renewcommand{\citenumfont}[1]{S#1}

Consider an object with the electric field distribution in the object plane given by $E(\vec\rho)$. Imaging the object with the objective lens will result in the electric field distribution in the image plane that is the convolution of $E$ and the point spread function $\psi$: 
\begin{align}
    \tilde E(\vec r) = \int E(\vec\rho) \psi(\vec r-\vec\rho) \de \vec\rho.
\end{align}
Heterodyne detection of the image field with the local oscillator prepared in a Hermite-Gaussian mode $\phi_{mn}(\vec r)$ will produce the photocurrent 
\begin{align}
    J_{mn}=\int\phi_{mn}(\vec r)\tilde E(\vec r) = \iint E(\vec\rho) \psi(\vec r-\vec\rho)\phi_{mn}(\vec r) \de \vec\rho.
\end{align}
 For incoherent objects, we measure the time-averaged power (absolute square) of the heterodyne photocurrent, which is given by 
\begin{align}
    P_{mn} = \langle |J_{mn}|^2 \rangle  &=  \iiiint \langle E(\vec\rho) E^*(\vec\rho') \rangle \psi(\vec\rho-\vec r) \psi^*(\vec \rho'-\vec r')  \phi_{mn}^*(\vec r) \phi_{mn}(\vec r') \ \de\vec \rho \de\vec \rho'\de\vec r\de\vec r'.
\end{align}
In the above result, we can recognise $\langle E(\vec \rho) E^*(\vec \rho') \rangle$ as the mutual intensity function. For a fully incoherent object with the intensity distribution $I(\vec\rho)$, the mutual intensity is 
 $$\langle E(\vec \rho) E^*(\vec \rho') \rangle=I(\vec\rho)\delta(\vec\rho-\vec\rho').$$
 Hence \begin{align}
    P_{mn} = \iiint I(\vec \rho) \psi(\vec \rho-\vec r) \psi^*(\vec \rho-\vec r') \phi_{mn}^*(\vec r) \phi_{mn}(\vec r') \ \de\vec \rho  \de \vec r\de\vec r' .
\end{align}
The above integral can be written as 
\begin{align}
    P_{mn} = \int I(\vec \rho) \left|[\phi_{mn} * \psi](\vec\rho)\right|^2, 
\end{align}
where $\phi_{mn} * \psi$ is the convolution of
 the Hermite-Gaussian mode
$$\phi_{mn}(\vec r) = \frac{H_k(\frac{x}{\sqrt{2}\sigma}) H_k(\frac{y}{\sqrt{2}\sigma})}{\sqrt{2^{m+n}m!n!}} \psi(\vec r)$$
and the Gaussian PSF 
\begin{align}
    \psi(\vec\rho) = \frac{1}{\sqrt{2\pi\sigma^2}} e^{-\frac{x^2+y^2}{4\sigma^2}},
\end{align} 
which is assumed to match $\phi_{00}$. This convolution is known to equal \cite{Weierstrass} 
\begin{align}
    \left[\phi_{mn} * \psi\right](x,y) = \frac{1}{\sqrt{m!n!}} \left ( \frac{x}{2\sigma} \right )^m \left ( \frac{y}{2\sigma} \right )^n e^{-\frac{x^2+y^2}{8\sigma^2}},
\end{align}
so we finally get
\begin{align}\label{Pmn}
    P_{mn} &=  \frac{1}{m! n!} \int I(x,y) \left( \frac{x}{2\sigma} \right )^{2m} \left( \frac{y}{2\sigma} \right )^{2n} e^{-\frac{x^2+y^2}{4\sigma^2}} \ dx dy.
\end{align}
From these measurements, we can calculate the following sums with even indices $p$ and $q$,
\begin{align}
    \beta_{pq} = \sum_{\text{even }k,l=0}^{p,q} \left(\frac{k}{2}\right)!\left(\frac{l}{2}\right)! \alpha_{pk} \alpha_{ql} P_{\frac{k}{2}\frac{l}{2}},
\end{align}
with the $\alpha$'s being the coefficients of Hermite polynomials $H_i(x) =  \sum_j^i \alpha_{ij} x^j$. 
These sums are equal to scalar products between the object intensity distribution and the corresponding Hermite-Gaussian polynomials:
\begin{align}
    \beta_{pq} &= \sum_{\text{even }k,l=0}^{p,q} \alpha_{pk} \alpha_{ql}  \int I(x,y) \left( \frac{x}{2\sigma} \right )^{k} \left( \frac{y}{2\sigma} \right )^{l} e^{-\frac{x^2+y^2}{4\sigma^2}} \ dx dy\\
    &= \sum_{\text{even }k,l=0}^{p,q}  \int I(x,y) \phi_{kl}(x,y) \ dx dy.
\end{align}

Because 2-dimensional Hermite-Gaussian polynomials $\phi_{pq}$ are orthonormal with weight $2^{p+q+2}p!q!\pi\sigma^2$, we have that $\beta$'s are the expansion coefficients of $I(x,y)$ into the Hermite-Gaussian basis:
\begin{align}
    I_{\rm even}(x,y) = \sum_{p,q = 0}^N \frac{\beta_{pq} \phi_{pq}(x,y)}{2^{p+q+2}p!q!\pi\sigma^2},
\end{align}
Because the measurements only recover the $\beta$'s with even indices, we can reconstruct the even component of the object intensity distribution $I(x,y)$. In other words, if we calculate Eq.~(1) from the main text including only even $m$ and $n$, we will reconstruct the distribution $[I(x,y)+I(-x,y)+I(x,-y)+I(-x,-y)]/4$.

Alternatively, we can recover the odd components by measuring in the overcomplete measurement basis, which includes HG mode superpositions
$$\phi_{mn}^{0-6}:=\{\phi_{m,n}, \phi^\pm_{m,m+1, n,n}, \phi^\pm_{m,m,n,n+1}, \phi^\pm_{m,m+1,n,n+1} \},$$
where
\begin{align}
    \phi^\pm_{m,m',n,n'} := \frac{1}{\sqrt{2}} [\phi_{m',n'} \pm \phi_{mn}].
\end{align}
By measuring the heterodyne photocurrent powers $P_{mn}^{(0)-(6)}$ corresponding to these mode superpositions, we can compute the integrals 
\begin{align}\label{PmnPrime}
    Q_{mn} &=  \int I(x,y) \left( \frac{x}{2\sigma} \right )^{m} \left( \frac{y}{2\sigma} \right )^{n} e^{-\frac{x^2+y^2}{4\sigma^2}} \ dx dy.
\end{align}
according to 
\begin{align} 
Q_{mn} &= 
\begin{cases}
    \left ( \frac{m}{2} \right )! \left ( \frac{n}{2} \right )! P^{(0)}_{\frac{m}{2}\frac{n}{2}} & m,n\text{ even}\\
    \left ( \frac{m-1}{2} \right )! \left ( \frac{n}{2} \right )! \sqrt{\frac{m+1}{8}} \left ( P^{(1)}_{\frac{m-1}{2}\frac{n}{2}} - P^{(2)}_{\frac{m-1}{2}\frac{n}{2}} \right ) & \text{$m$ odd, $n$ even}\\
    \left ( \frac{m}{2} \right )! \left ( \frac{n-1}{2} \right )! \sqrt{\frac{n+1}{8}} \left ( P^{(3)}_{\frac{m}{2}\frac{n-1}{2}} - P^{(4)}_{\frac{m}{2}\frac{n-1}{2}} \right ) & \text{$m$ even, $n$ odd}\\   
    \left ( \frac{m-1}{2} \right )! \left ( \frac{n-1}{2} \right )! \sqrt{\frac{m+1}{8}} \sqrt{\frac{n+1}{8}} \left ( P^{(5)}_{\frac{m-1}{2}\frac{n-1}{2}} - P^{(6)}_{\frac{m-1}{2}\frac{n-1}{2}} \right ) & m,n\text{ odd.}
\end{cases}
\end{align}
Knowing $Q_{mn}$, we can evaluate 
\begin{align}
    \beta_{pq} = \sum_{m,n=0}^{pq}  \alpha_{pm} \alpha_{qn} Q_{mn}
\end{align}
for all $p$ and $q$, and hence reconstruct $I(x,y)$.

\end{document}